\documentclass[11pt,preprint]{aastex}

\newcommand{\etal}{{\it et~al.}}

\usepackage{longtable}

\begin{document}

\title{Validation of the Survey Simulator tool for the NEO Surveyor mission using NEOWISE data}

\author{Joseph R. Masiero\altaffilmark{1}, Dar W. Dahlen\altaffilmark{1}, Amy K. Mainzer\altaffilmark{2}, William F. Bottke\altaffilmark{3},  Jennifer C. Bragg\altaffilmark{2}, James. M. Bauer\altaffilmark{4}, Tommy Grav\altaffilmark{2}}

\altaffiltext{1}{Caltech/IPAC, 1200 E. California Blvd, MC 100-22, Pasadena, CA 91125 USA}
\altaffiltext{2}{University of Arizona, Tucson, AZ 85721 USA}
\altaffiltext{3}{Southwest Research Institute, Boulder, CO 80302 USA}
\altaffiltext{4}{University of Maryland, College Park, MD}

\begin{abstract}

  The Near Earth Object Surveyor mission has a requirement to find
  two-thirds of the potentially hazardous asteroids larger than 140
  meters in size.  In order to determine the mission's expected
  progress toward this goal during design and testing, as well as the
  actual progress during the survey, a simulation tool has been
  developed to act as a consistent and quantifiable yardstick.  We
  test that the survey simulation software is correctly predicting
  on-sky positions and thermal infrared fluxes by using it to
  reproduce the published measurements of asteroids from the NEOWISE
  mission. We then extended this work to find previously unreported
  detections of known near Earth asteroids in the NEOWISE data
  archive, a search that resulted in 21,661 recovery detections,
  including 1,166 objects that had no previously reported NEOWISE
  observations.  These efforts demonstrate the reliability of the NEOS
  Survey Simulator tool, and the perennial value of searchable image
  and source catalog archives for extending our knowledge of the small
  bodies of the Solar System.

\end{abstract}

\section{Introduction}

The Near Earth Object Surveyor mission is being designed and built
with the explicit requirement of detecting two-thirds
  of all large, close-approaching near Earth objects (NEOs), and has
the extended goal of fulfilling the George E. Brown
Act\footnote{https://www.congress.gov/109/plaws/publ155/PLAW-109publ155.pdf}
which requires NASA to find and assess the hazard to Earth of over
$90\%$ of the potentially hazardous asteroids larger than 140 meters
in
diameter \footnote{https://www.nasa.gov/pdf/171331main\_NEO\_report\_march07.pdf}.
To do this the project will place a passively-cooled, thermal infrared
telescope at the Sun-Earth L1 point. This will allow for a survey to
be conducted at Solar elongations down to $45^\circ$, a region of sky
that is difficult to survey with ground-based telescopes but where
NEOs on orbits like the Earth's will spend a much larger
  fraction of time compared to the opposition region.  For a more
detailed description of NEO Surveyor please refer to
\citet{mainzer23}.

As part of the analysis conducted by \citet{mainzer23}, a simulation
of the planned NEO Surveyor observing pattern was carried out to
determine the completeness for NEOs that will be attained (that is,
the fraction of the input population that would be detected by the
survey).  This simulation used a synthetic population of small Solar
System bodies and the planned observing sequence over the 5-year
survey period to determine when objects would be detected and if the
detections would be sufficient for an object to be cataloged as
discovered.  The outputs of this NEO Survey Simulator (NSS) will be
used by the project to assess the relative influence of different
parameters within the Observatory design and its concept of operations
on the final completeness, the top-level scientific margin being held
by project during development, and the completeness delivered by the
mission during survey operations.

\citet{mainzer23} describe the validation of the synthetic Solar
System model used as input for our survey simulations.  In this work,
we describe efforts to validate the critical software components of
the NSS. In particular we focus on ensuring that the NSS is correctly
determining the positions of the synthetic objects, that the
geometry-checking routines that determine if the asteroid is in the
telescope's field of view are correct, and that the predicted thermal
fluxes created by the NSS match the actual expected fluxes.  To do
this, we make use of data from the Near Earth Object Wide field
Infrared Survey Explorer \citep[NEOWISE;][]{mainzer11,mainzer14neowise}
which measured thermal infrared fluxes for over $150,000$ asteroids in
2010 during its cryogenic phase and continuing to today in the
Reactivation survey has provided thermal IR measurements for nearly
$2000$ near Earth asteroids.  NEOWISE provides an ideal dataset to
test the NEO Surveyor tools and validate the NSS.

\section{Survey Simulator Design}

As discussed in \citet{mainzer23}, the NEO Surveyor mission has the
top-level requirement of detecting and cataloging at least two-thirds
of all asteroids larger than 140~m in size and that approach within
$0.05~$AU of the Earth's orbit (i.e. Minimum Orbit
  Intersection Distance MOID$<0.05~$AU), known as the Potentially
Hazardous Asteroids (PHAs).  This requires a system with a sufficient
sensitivity to measure the thermal emission of these objects, a
sufficient sky coverage to sample the whole population, a sufficient
return time to obtain a series of detections of an object (known as a
``tracklet'') and link these tracklets together to obtain an orbit,
and a sufficient survey duration to sweep up objects with long orbital
periods or unusually long synodic periods.  Each of these needs drives
an aspect of the mission design.  This design, however, allows for a
range of possible concepts of operations for the survey, each with its
own expected final output.  To evaluate the feasibility and value of
potential trade-offs and ensure that the chosen concept meets the
mission Level 1 goals, a survey simulator has been built that computes
the expected catalog completeness of a reference population using the
mission design and survey operations parameters for different test
cases.

The NSS takes as input three primary data sources.  First, a database
containing a population of objects including orbital and physical
properties is required in order to evaluate the potential for
detectability of each object.  Second, a plan for the individual
telescope pointings is required.  Finally, the properties of the
observatory (such as sensitivity, field of view, and wavelength, etc.)
are required.  With these data as inputs, it is possible to simulate
the planned survey to evaluate if each object is in the field of view,
is detectable, and has sufficient detections to be cataloged. The
principal output of the NSS is a plot of the catalog completeness of
the target population as a function of time.

A top-level description of the simulation steps followed by the NSS are provided as follows, with a more detailed discussion given in \citet{mainzer23}:
\begin{enumerate}

\item For each observation in the survey plan, the Ecliptic state vectors
  for each object in the input population are propagated to the time
  of observation.  Object positions are corrected for light-time delay
  between the object and spacecraft, and then each object is evaluated
  to check if it fell within the footprint of the active area of a
  detector.  The detector sizes and layouts are defined in the input
  configuration file.  This step results in a list of potential
  detections, i.e. those objects that are geometrically accessible.
  
\item For each potential detection of an object, the object's physical properties
  are used along with the observing geometry, to calculate the expected flux from the
  object at each NEO Surveyor bandpass.  The flux is a combination of
  reflected light following the predictions from the H-G formalism
  \citep{bowell89} and thermal emission using either the Near Earth
  Asteroid Thermal Model \citep[NEATM,][]{harris98} or the Fast
  Rotating Model \citep{lebofsky78} (as specified in the configuration
  file).

\item The calculated flux for each observation is compared to the
  sensitivity of the detectors for the given observing geometry. This
  sensitivity will not be constant as the flux from the zodiacal
  background changes dramatically based on wavelength and the relative
  positions of the Sun, ecliptic plane, and telescope field of view.
  The predicted sensitivity as a function of viewing geometry relative
  to the Sun (accounting for both the detector performance and
  zodiacal background) is included with the NSS as a data file lookup
  table.  The expected flux from the zodiacal background is based on
  the \citet{wright05} model of the zodiacal dust cloud.  Two versions
  of the sensitivity file are included with the NSS code: one
  corresponding to the Current Best Estimate (CBE) of system
  performance, and one set to the mission requirements to represent
  the limiting case.

\item The NSS then determines if the survey would build a tracklet for
  an object based on the signal-to-noise cutoff for a detection, the
  tracklet assembly requirements, tracklet velocity limits, and the
  detection and tracklet efficiency measured from data processing
  simulations at the NEO Surveyor Survey Data Center (NSDC).
  In the nominal case, a tracklet is assumed to be built if
    the object is detected 4 or more times at an SNR$>5$ and has an
    on-sky motion larger than 0.008 deg/day and smaller than 8
    deg/day. Tracklet building efficiency has been measured at $99\%$
    from recent simulation tests, and so $\sim1\%$ of tracklets that
    pass the above thresholds are dropped randomly to simulate this
    incompleteness.

\item Tracklets are assembled into tracks based on the Minor Planet
  Center's historical and simulated efficiency of linking isolated
  tracklets into tracks that are sufficient to compute accurate orbits
  and thus be cataloged as NEOs.  In the nominal case, we
    assume a linking efficiency of $99\%$ based on historical
    performance of the MPC from NEOWISE survey data; future testing
    will provide a constraint on this parameter for the expected
    Surveyor observing cadence.
  
\item Using the simulated survey parameters and the model
  population, the NSS calculates the completeness that would be
  expected to be achieved as a function of time. The survey
  completeness is calculated as the fraction of objects in the input
  population that have sufficient data to be recorded as
  tracks by the Minor Planet Center (MPC) and have orbits determined.
  This fraction as a function of survey duration is used to
  evaluate the overall expected completeness of the survey, the
  effects of any changes to the survey plan or system configuration,
  and the scientific margin that the mission is carrying.  In
    this way, the NSS is able to provide verification that the planned
    survey would fulfill the mission Level 1 requirements.

\end{enumerate}

\citet{mainzer23} describe the anticipated survey completeness for NEO
Surveyor based on the current mission design, as well as some of the
considerations that drove design decisions.  In order to be confident
in the results of these studies, it is necessary to demonstrate that
the predictions being made by the NSS, particularly the on-sky
position and flux predictions, are accurate.  This is doubly important
as the outputs of the NSS are used by the NSDC as inputs to the
ongoing image simulation work that is allowing the mission to develop
and test the data reduction pipeline prior to the launch of the
mission.  In the following sections we describe our methods for
validating the NSS outputs.

\section{Comparison to Horizons predictions}

Our first validation of the NSS seeks to confirm that the positions of
asteroids are being correctly predicted.  In order to demonstrate
this, we must identify a source of `truth' values that our outputs can
be checked against. As a first test of the NSS, we elect to use the
JPL Horizons
tool\footnote{\textit{https://ssd.jpl.nasa.gov/horizons/app.html}} to
provide comparison values for the astrometric positions of our
simulated objects.  This test was carried out for both real asteroids
using their published orbital element information as well as for
synthetic objects computed in parallel by Horizons and by the NSS
tools.

\begin{figure}[ht]
\begin{center}
  \includegraphics[scale=0.35]{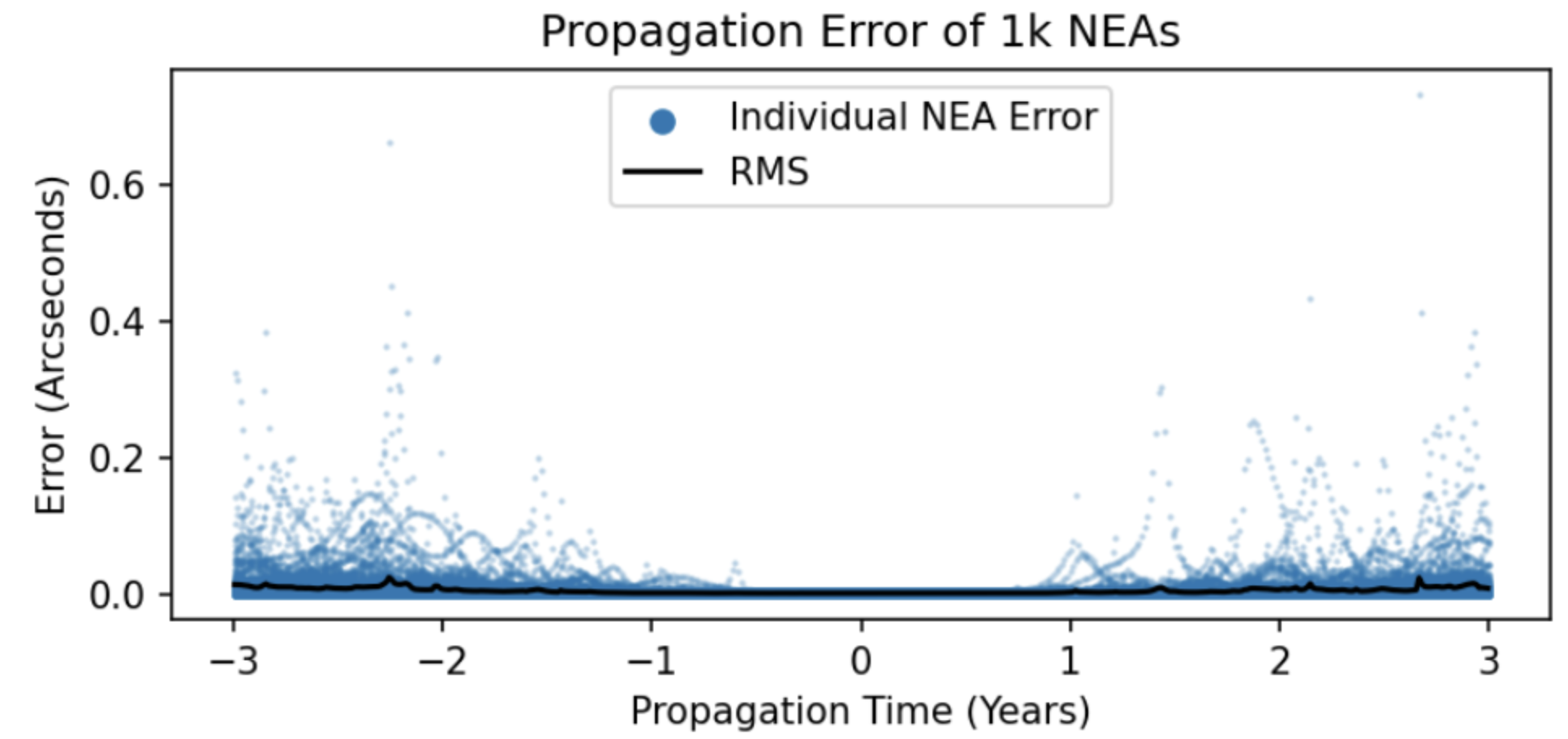} \protect\caption{
    Comparison of NSS-predicted positions for 1000 known NEOs with
    well-determined orbits (i.e. numbered asteroids) to the JPL
    Horizons position of each object.  The errors for each object are
    plotted as a blue point at each timestep evaluated. 
      Errors for individual objects will grow well above the RMS
      value, and then decrease in cases where an object has a close
      pass with the telescope. The black line shows the RMS error for
    the population evaluated at each timestep.}
\label{fig.known_astrom}
\end{center}
\end{figure}

The NSS uses as a starting point an osculating orbital element set
obtained from the MPC for each object, and performs an N-body
gravitational orbital evolution from that point to the time of
observation, accounting for first- and second-order relativistic
terms.  The Sun, Moon, and planets are all used as massive
  bodies for this calculation; the massive asteroids are neglected
  given the short time span required for this work.
\citet{mainzer23} provide a detailed description of the method of
implementation for these effects.  Positions are computed in this way
at hourly timesteps.  For times between these benchmarks, including
correcting for light-time delays, the nearest Cartesian state vector
and velocity are propagated linearly in three dimensional space.
Figure~\ref{fig.known_astrom} shows the offset between the
NSS-determined astrometric position and the JPL Horizons position for
1000 known NEOs with good orbits (i.e. numbered asteroids with over 3
years of orbital arc).  For this test, the position of the objects
were defined at the center time of the simulation and propagated
forward and backward in time for a sufficient period to cover the
entire NEO Surveyor survey.  This reduces the numerical errors that
build up over time and is the same process that is being used during
survey simulation and completeness determination, with the epoch of
synthetic object defined at the center time of the survey.

As shown in Fig~\ref{fig.known_astrom}, the NSS code produces
positions with an on-sky RMS accuracy better than the mission
requirement value of an RMS of $0.1$ arcsec in each axis.  Objects
with larger offsets are traced to those NEOs that have very low
perihelia, resulting in a build up of numerical noise that increases
the on-sky offset at times of close passes with the Earth.  The on-sky
positional errors subsequently decrease as the object recedes from
Earth, and are in all cases less than 1 arcsec.  This demonstrates that
the N-body propagation code being used by the NSS accurately reflects
both Horizons and the data we expect to obtain when NEO Surveyor
begins collecting data.

\section{Comparison to published NEOWISE data}

While comparisons of one computer model to another offer an important
confirmation that the NSS is correctly computing expected parameters,
comparisons to measured data offer an independent means of
confirmations of the validity of the methodology.  In addition,
comparisons to Horizons do not offer the ability to validate the
thermal emission model used in the NSS.  To address the ability to
correctly compute both the ephemerides as well as fluxes, we next
performed a validation of the NSS by comparing our predicted results
to the reported NEOWISE astrometric and thermal flux measurements
that are obtained from the IRSA data
  archive\footnote{https://irsa.ipac.caltech.edu/applications/Gator/}.

The NEOWISE mission obtained thermal infrared observations of over
150,000 asteroids and comets over the course of its multiple mission
phases.  The original NEOWISE data \citep{mainzer11} were obtained as
part of the WISE survey \citep{wright10} from 7 January 2010 to 6
August 2010, and simultaneously observed at $3.4~\mu$m, $4.6~\mu$m,
$12~\mu$m and $22~\mu$m.  After the exhaustion of the outer cryogen
tank the survey continued in 3-band cryogenic mode though 29 Sep 2010,
and then post-cryogenic mode until WISE was put into hibernation on 1
Feb 2011 \citep{cutri12}.

NEOWISE reported observations for $428$ NEOs during the fully
cryogenic phase of the mission \citep[i.e. when all four channels were
  operational;][]{mainzer11neo}.  While this population spans the
range of magnitudes that we wish to validate here, we desire a larger
sample to better investigate any systematics in our flux calculation.
To that end, we also consider the $128,462$ Main Belt asteroids
that were detected by WISE during the fully cryogenic mission.  This
population of known objects is used to validate our astrometric
computation from a space-based observatory and our calculations of
thermal flux.

To check the NSS astrometric accuracy, we downloaded from IRSA the
time and pointing of every WISE frameset from the fully cryogenic
phase of the mission.  We also queried Horizons for the position and
velocity of the WISE spacecraft at 15-minute increments over the same
time span.  These were used in place of the survey pattern and
spacecraft ephemerides for NEO Surveyor.  Using initial orbits from
Horizons for the objects known to have been detected by NEOWISE, we
built a model population and used the NSS tools to compute the
predicted positions and fluxes for each object.

\begin{figure}[ht]
\begin{center}
  \includegraphics[scale=0.7]{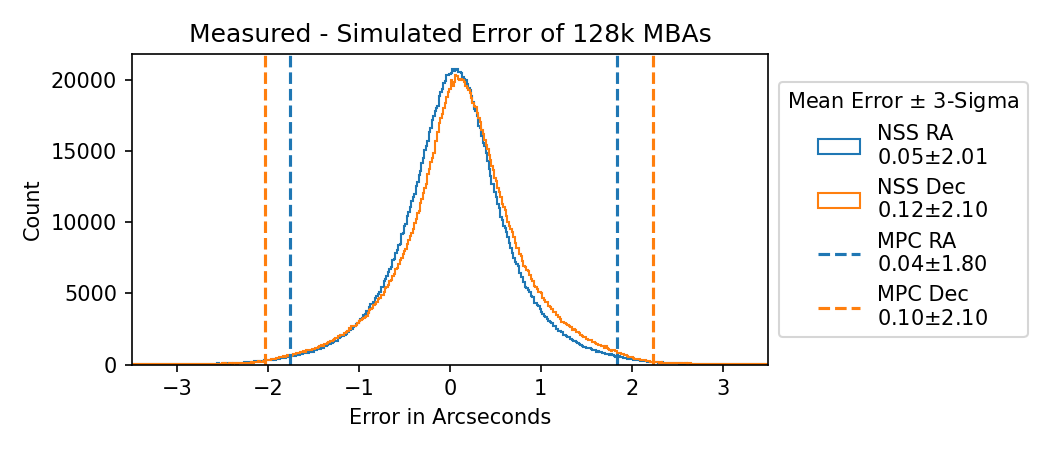}
  \protect\caption{ Histogram showing the comparison of NSS-predicted
    astrometric positions in Right Ascension (blue) and Declination
    (orange) to the measurements reported to the MPC for $1,624,795$
    detections of $128,462$ MBAs observed by NEOWISE during the 2010
    fully cryogenic mission.  The vertical dashed lines show the
    offset and $3-\sigma$ scatter provided by MPC's analysis of
    observations residuals for the submitted C51 observations to
    predictions from orbits of known objects.  The offsets and random
    scatter in the observation error from the NSS code are comparable
    to the values found by the MPC's analysis.}
\label{fig.nw_astrom}
\end{center}
\end{figure}

We compared the positions predicted by the NSS to the measurements
reported by the NEOWISE mission to the MPC.  We find that every 
detection that was reported to the MPC was listed as a potential
detection by the NSS.  Figure~\ref{fig.nw_astrom} shows the offsets
between $1,624,795$ predicted and measured on-sky positions for $128,462$
Main Belt asteroids detected in this period.  The Main Belt is used
here as there are over two orders of magnitude more objects detected
than for the NEOs, and it therefore provides superior statistics for
constraining the accuracy of our simulation.

The systematic offsets found in this test ($0.05''$ and $0.12''$ for
RA and Dec respectively) are significantly less than the $2.75''$ size
of a WISE pixel, and the scatter ($2.01''$ and $2.10''$ 3-sigma in RA
and Dec, respectively) is much less than the $7.25''$ PSF width.  The
offsets and scatter also closely match the astrometric residuals for
the WISE cryogenic mission recorded by the MPC\footnote{Systematic
offsets and 1-sigma scatter for 2010 for NEOWISE observatory code C51
are given in the MPC's analysis of observation residuals available
here:
\textit{https://minorplanetcenter.net/iau/special/residuals.txt}}.
These results further demonstrate that the orbit propagation and
positional determination in a spacecraft's field of view has been
implemented correctly and with sufficient precision for the needs of
the NEOS project.

The vast majority of objects detected by NEOWISE during the cryogenic
phase of the mission had sufficient data to fit a thermal model, and
thus have physical properties reported
\citep[e.g.][]{mainzer11neo,masiero11,mainzer19}.  We take the
physical properties for each epoch of observation and use them in the
NSS to create a predicted thermal infrared flux at the time of
observation.  The sensitivity, zero point, and central
  wavelength for each band are taken from the WISE Explanatory
  Supplement \citep{cutri12}.  The predicted value is then compared to
the magnitude published for that observation in the NEOWISE data
archive in IRSA.  This comparison allows us to validate that our
thermal modeling code is correctly implemented.

\begin{figure}[ht]
\begin{center}
  \includegraphics[scale=0.5]{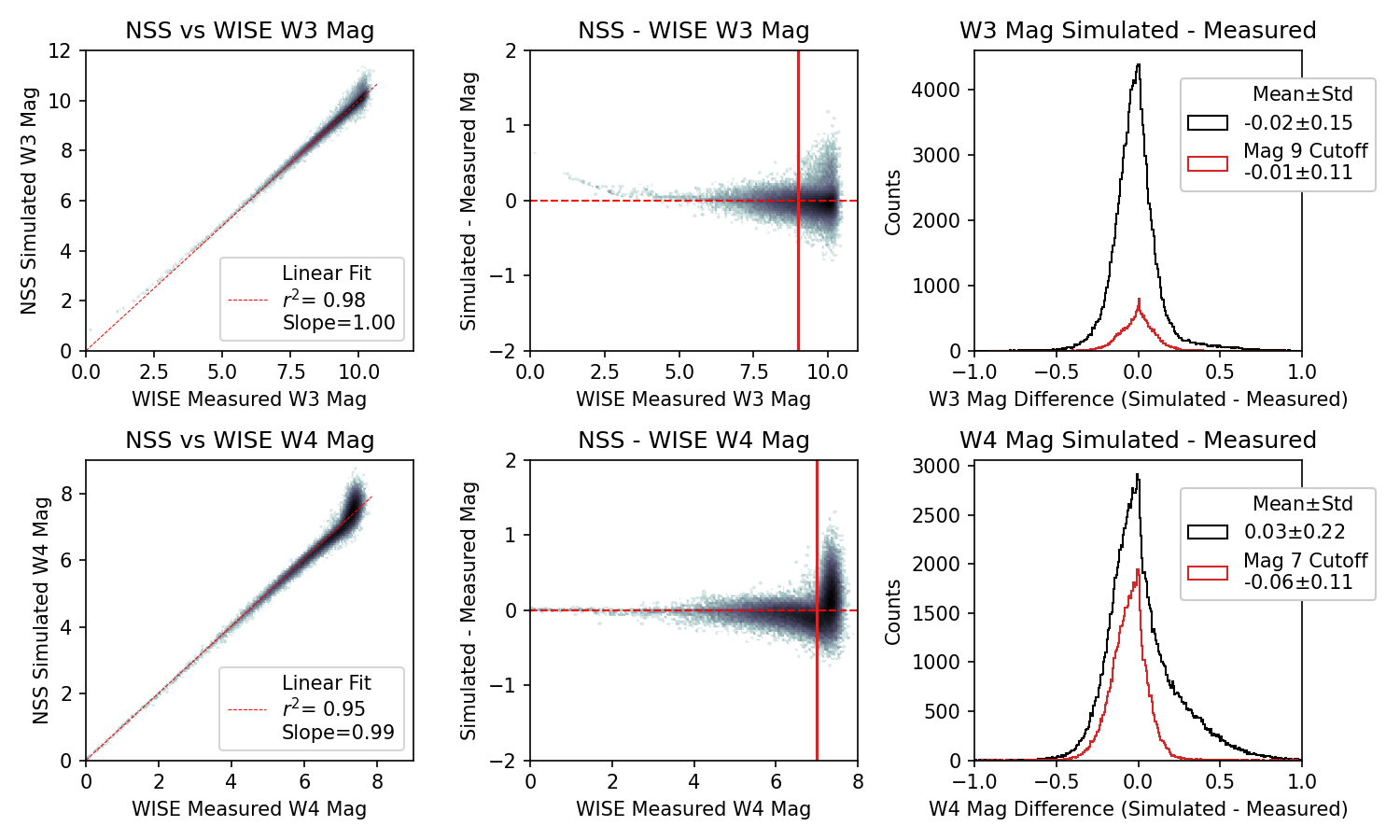}
  \protect\caption{Validation of the NSS thermal model is done through
    comparison of the predicted magnitude to the WISE-measured
    magnitudes for W3 (top) and W4 (bottom). Columns show: (left) a
    density plot of the comparison between the prediction and the
    measurement in grey along with a best-fit linear trend in red;
    (middle) a density plot of the measured magnitude against the
    difference between the simulation and the measurement with the
    vertical red line showing the approximate transition into a
    background-impacted regime; (right) histograms of the magnitude
    differences for all objects in black and those brighter than the
    background cutoff in red. The spatial bins in the density plots
    are shaded following a logarithmic scale to emphasize the bright
    tail.}
\label{fig.nw_mags}
\end{center}
\end{figure}

The NEOWISE physical properties were determined by fitting the NEATM
thermal model to all detections in a given observing epoch
\citep{mainzer11neatm}.  The result of this is that the best-fit
diameter represents a time-averaged, spherical equivalent size.  To
properly compare to the observations, we take the average predicted
magnitude over each individual observing epoch and compare that to the
average of the measured magnitudes at the same epoch.  The NEATM
beaming parameter used to calculate the flux at each epoch was drawn
from the \citet{mainzer19} data table.  Figure~\ref{fig.nw_mags} shows
the comparison of the predicted and measured magnitudes for the W3
($12~\mu$m) and W4 ($22~\mu$m) bands for $1,624,795$ detections of
$128,462$ Main Belt asteroids, as these two bands are always thermally
dominated for these objects.  The shorter wavelength channels can
include significant contributions from reflected light depending on
object temperature and involve more model-dependent assumptions about
the reflected light behavior at these wavelengths and the fractional
contributions of each.  As above, the Main Belt population was used
here to provide robust statistics over the full range of observed
magnitudes.

Our analysis shows that the predicted magnitudes generally match the
observed values to within the quoted uncertainties.  Given
  that the physical parameters used here were originally derived from
  WISE data, this demonstrates that the implementation of the model
  used here is consistent with that used to derive the physical
  properties. Two exceptions stand out: at the bright end a
systematic deviation is seen that is due to incomplete correction of
non-linearity and saturation effects in the simulated data, while at
the faint end the effects of background noise contributions to the
measured data are apparent.  This noise effect is especially
pronounced in W4, where the multi-band forced-photometry carried out
on the WISE images \citep[see][for details]{cutri12} causes measured
values to become brighter than the prediction.  This effect
  occurs because asteroids tend to be brightest in W3, and so a faint
  W3 source will report a measurement for W4 that is
  background-dominated and so preferentially brighter than expected.
In the regime above the faint limit, the NSS prediction matches the
observations, confirming that our implementation of the predicted
thermal emission is correct, with a scatter of only $\sim0.11~$mag and
systematic offsets well below $0.1~$mag.

\clearpage

\section{Recovery of new NEOWISE detections of known NEOs}

Having demonstrated that the NSS can successfully reproduce the
positions and magnitudes of the asteroids already measured by NEOWISE,
it becomes natural to ask if this tool can identify detections in the
NEOWISE data that have not yet been reported.  Searches of NEOWISE
data for unreported detections of known NEOs have been successfully
conducted in the past \citep[see ][for
  details]{mainzer14tinyneo,masiero18tiny2,masiero20tiny3}.  However
there is a continuing benefit in revised searches of older data as new
objects are discovered and known objects receive improved orbital
solutions.  The new NSS software developed for the NEO Surveyor
provides the tools needed to quickly search the entire NEOWISE archive
for predicted detections of all known NEOs.

The population we used for our search was drawn from the Minor Planet
Center's MPCORB
file\footnote{\textit{https://minorplanetcenter.net/iau/MPCORB.html}},
filtered to only include those objects with perihelion distances of
$q<1.3~$AU.  This query was carried out on 17 Aug 2022, and resulted
in a list of 29231 objects. Five objects that are listed in the MPCORB
catalog but hit the Earth shortly after discovery were filtered from
the search; as all of these objects have $H_V>29~$mag they are almost
certainly too small to have been detected by NEOWISE.  Objects
  with short observational arcs were not specifically removed, so that
  NEOWISE detections close in time to the observed arc might be
  recovered. These objects will have large positional uncertainties at
  other times, but visual vetting of recovered detections will remove
  spurious associations.

For objects where the physical properties were known, we used the
measured diameter and albedos for the search.  For all other objects
we assumed an albedo of $p_V=5\%$ and derived a diameter based on the
published $H_V$ magnitude which will result in the objects being
preferentially larger than reality and make the fluxes predicted by
the NSS larger.  This ensures that our list of potential detections is
as comprehensive as possible.

As was done for the test described above, we used the structure of the
NSS code to simulate the position and viewing geometry of NEOWISE.
The IRSA Single-Exposure Frame Metadata tables for each mission phase
were queried without constraint to determine the position of the field
of view of each frame with the associated MJD.  These then became the
``Visits'' that would be used by the NSS.  The frame metadata tables
do not include the spacecraft position, so to calculate that we
downloaded from Horizons the position of the WISE spacecraft at
15-minute intervals for the duration of all phases of the mission from
07 Jan 2010 to 13 Dec 2021.  These way-points were propagated in time
to the center point of each exposure to get the spacecraft position at
each Visit.

The state vector for each object was propagated to the time of each
exposure, and compared to the field of view of the detectors.  Objects
that fell on at least one detector were considered Potential
Detections.  To determine detectability, we convert the magnitude in
each band for a signal-to-noise ratio (SNR) of five given in the WISE
and NEOWISE Explanatory Supplements \citep{cutri12,cutri15} into fluxes,
and compare those values to the flux determined for the object using
NEATM.  All objects passing this threshold were reported by the NSS
($\sim30,000$ instances per year for the two band data and
$\sim37,000$ instances in the eight months of 4-band and 3-band cryo
mission phases).

This list should then include all possible detections of all currently
known NEOs during the NEOWISE survey.  This output was filtered to
remove all detections that had already been reported to the MPC,
either those found by the WMOPS automated processing or recovered in
previous searched of missed detections ($\sim11,000$ during the cryo
mission phases and $\sim4,000$ per year for the two-band data).
Taking the remaining list of potential detections that were not
present in the MPC observation file, we then searched the IRSA archive
of each mission phase at the predicted time and position, within 5
arcsecs of the predicted position and 5 seconds of exposure time of
the frame.  Approximately $70\%$ of potential detections had
  no associated source returned by IRSA; the most likely cause for
  this is that real NEOs had higher albedos than our assumed value
  which would translate to smaller sizes and fainter IR fluxes when
  using the $H_V$ magnitude and an assumed size.  This fraction is
  consistent with the known bias toward higher albedo NEOs for objects
  discovered by ground-based surveys \citep{masiero20tiny3}.

The data table files returned by IRSA were cleaned to remove sources
of contamination.  This included removing sources that were associated
with background stars; removing detections that have $rchi2$ data
columns larger than three, an indication from the PSF-fit quality
measurements that they are likely cosmic rays; and removing detections
that had colors consistent with being stellar in nature even if not
identified as stars \citep[in this case where $W1-W2<1$, cf.][for
  details]{masiero17}.  At the end of this cleaning process there were
2553 detections from the cryogenic mission phase, 817 from the 3-band
cryo phase, 1523 from the post-cryo survey, and $2000-2500$ detections
from each year of the NEOWISE Reactivation mission.

For each of the $22,842$ detections left after filtering, we generated
a finder chart showing the predicted position overlaid on a cutout of
each of the available bands.  Every detection was checked by-eye to
verify that the source looked reliable and that the measurement was
not contaminated by background objects, detector effects, or scattered
light.  An additional requirement was levied on instances of a single
predicted detection for an object, where we required that these cases
have a SNR$>5$ detection in at least two bands to help ensure that we were
detecting real sources.  In total, after by-eye filtering, we recover
$21,661$ detections of near Earth asteroids that were previously not
reported to the MPC.  When combined with all previous NEO detections
reported to the MPC we now have NEOWISE observations of 3330 NEOs,
representing over $10\%$ of all known NEOs.

Detections were broken up into epochs of observation, defined as
coherent sets of observations with at least three days of gap between
the last detection of one epoch and the first detection of the next.
A small number of objects had on-sky motions comparable to the
progression of the survey field of regard, resulting in hundreds of
detections spanning over a month.  In these cases epochs were defined
as lasting no longer than 10 days to minimize the changes of viewing
geometry during an epoch, and detections were split up to match this
requirement.

Of these $21,661$ new detections, $10,939$ are additional detections
of an epoch of observation that had already been reported to the MPC.
In these cases, either the detection fell just below the
signal-to-noise threshold used for tracklet construction in the
original WMOPS search, the detection was removed from consideration by
WMOPS due to proximity to a nearby inertial source, or the detection
links together other WMOPS-reported tracklets. This latter case most
often happened when an object was in the NEOWISE field of regard for a
long period of time and as such had significant curvature of motion
on-sky which resulted in WMOPS splitting the object into separate
tracklets.

\begin{figure}[ht]
\begin{center}
  \includegraphics[scale=0.75]{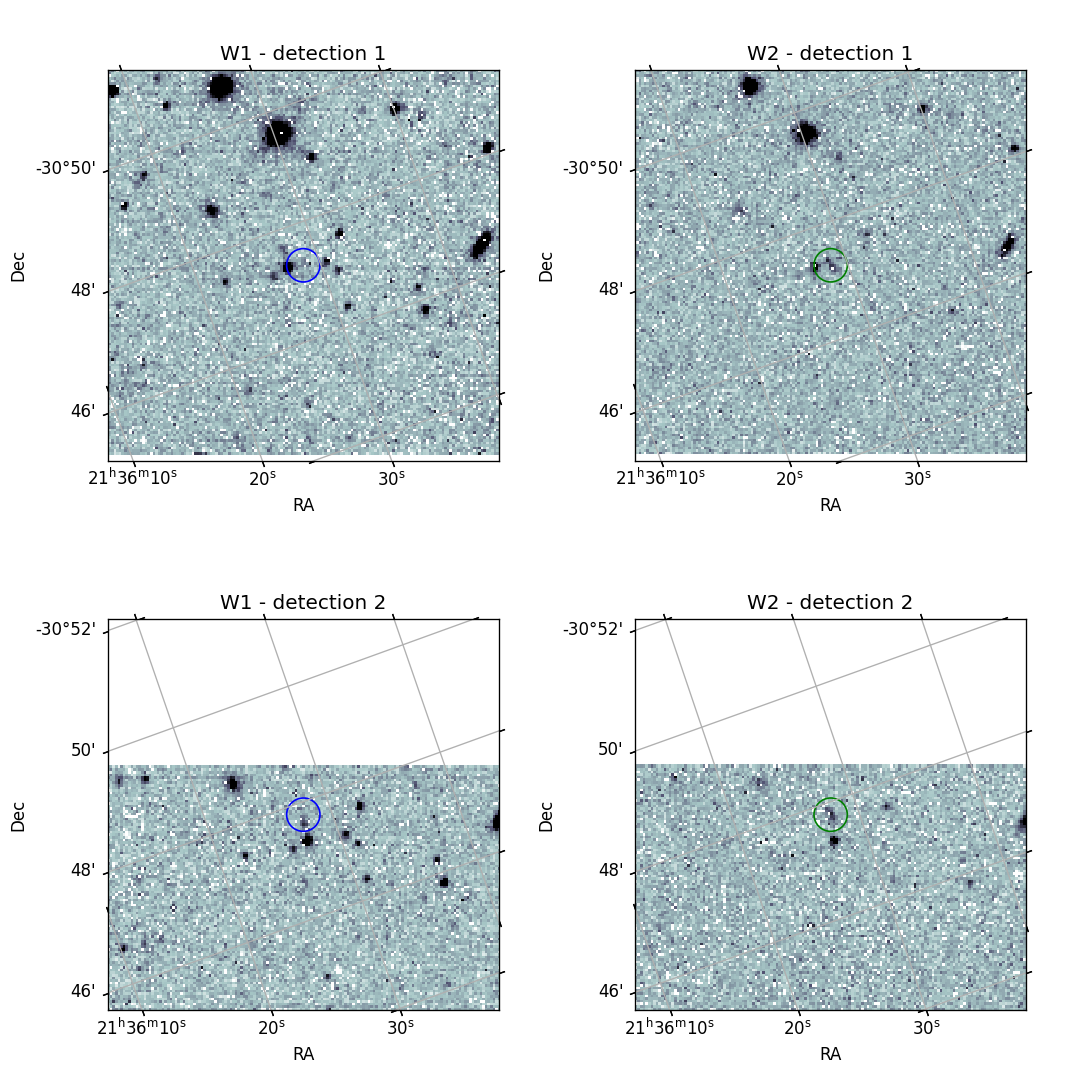}
  \protect\caption{Detections of Near Earth Object 2020 TK$_3$ in the
    NEOWISE-Reactivation Year 7 data that were identified by the
    NSS. This object was seen in two sequential image sets (top and
    bottom) with center times separated by only 11 seconds.  The left
    and right columns show the W1 and W2 images for each detection,
    respectively, and the NSS-predicted position for the object is
    marked by the colored circle overlaid on each image. The object was
    moving at $70.5~$ deg/day at the time of observation; it is
    noticeably trailed in the $7.7~$second-long W2 exposures and
    motion with respect to the background is apparent.  Detection 2 is
    near the edge of the field of view, resulting in the image being
    cut off in the figure.}
\label{fig.2020TK3}
\end{center}
\end{figure}

\begin{figure}[ht]
\begin{center}
  \includegraphics[scale=0.8]{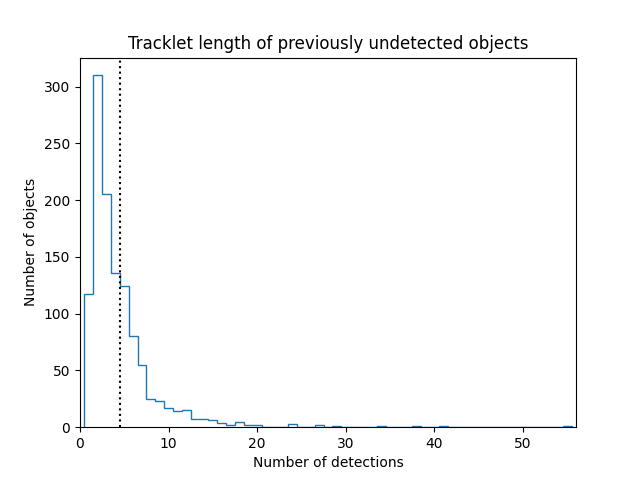}
  \protect\caption{Length of tracklets found in our search of NEOWISE
    data for objects that had no previously reported measurements.
    The majority have less than five detections (left of the
      dotted line) and could not have been found by the automated
    processing.  The longest tracklet here is 55 detections, for
    object 2017 SG14.}
\label{fig.newtracklet}
\end{center}
\end{figure}

The remaining recovered detections fall into two groups. The first
group contains $5,160$ detections of $1,166$ objects for which no
previous NEOWISE measurements had been reported to the MPC. Of these,
$768$ tracklets had fewer than five detections, and so would not be
found by WMOPS, which requires at least five detections to build a
tracklet.  An example of this is given in Figure~\ref{fig.2020TK3},
which shows two NEOWISE observations taken 11 second apart of the NEO
2020 TK$_3$ from the 7th year of the Reactivation Survey, which had an
on-sky rate of motion of $70.5~$deg/day at the time of observation.
$105$ of these objects with previously unreported astrometry ($355$
detections) were included in the physical property analysis conducted
by \citet{mainzer14tinyneo} but not reported to the MPC.
Figure~\ref{fig.newtracklet} shows the distribution of tracklet
lengths of these previously unreported objects.  The remaining $5,562$
detections represent new observation epochs of $786$ previously-reported
objects.

\clearpage

\begin{figure}[ht]
\begin{center}
  \includegraphics[scale=0.8]{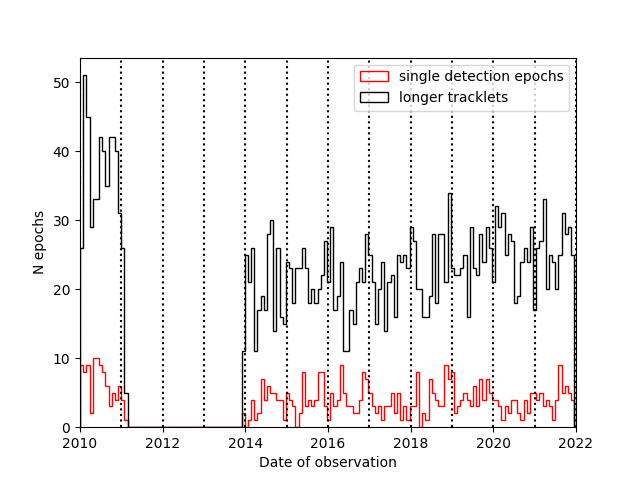} \protect\caption{
    Histogram showing the number of epochs of new detections recovered
    as a function of time.  The binsize used is 28 days, and the
    vertical dotted lines mark Jan 1 for each year starting with 2010.
    The red histogram shows epochs that consist of only a single
    detection, while the black histogram is all other recovered
    epochs.  The large gap from 2011-2014 is the time when
    the WISE spacecraft was in hibernation.  The recovery rate is
    higher during the cryogenic phase as the telescope was more
    sensitive then, and because the astrometry for the objects
    recovered by \citet{mainzer14tinyneo} was not submitted to the MPC
    and so does not appear in the list of already-reported
    detections. The general trend of increasing epochs with time
    during the reactivation mission is a result of the increasing
    number of objects discovered by ground-based surveys.  }
\label{fig.ydhist}
\end{center}
\end{figure}

We show in Fig~\ref{fig.ydhist} the count newly recovered epochs of
NEO detections as a function of time, splitting out objects detected
in only a single image with those seen in multiple images.  While the
detection count shows some structure, it overall is relatively flat.
The apparent structures seen in this histogram are:
\begin{itemize}
  \item The $\sim3-$year gap readily apparent is due to the period of
    time when the WISE spacecraft was placed in hibernation, from 1
    Feb 2011 until the survey resumed on 13 Dec 2013
    \citep{mainzer14neowise}.
  \item The 2010 period shows a larger number of recovered detections
    than later times; this is due to two primary effects: 1) WISE was
    more sensitive to asteroids during the cryogenic portions of the
    mission and 2) detections identified and used for thermal fitting
    by \citet{mainzer14tinyneo} were not reported to the MPC and so
    will be included in the new detection counts shown here.
  \item For the data collected after 2013, there appears to be a
    slight increasing trend in the number of recovered epochs with
    time. This is consistent with increase in NEO detection rate by
    ground-based surveys during this time, as the period around
    discovery is the most likely time for a close, bright pass with
    the Earth and thus a detection by NEOWISE.
\end{itemize}

\clearpage

The recovered NEO detections found in this search show no significant
trend with on-sky position.  The distribution of declinations follows
the expected $\cos\left({\rm Dec}\right)$ behavior and the
distribution of Right Ascensions is flat.  This holds for both the
single-detection epochs as well as the longer tracklet epochs.  The
recovered detections are also uniform with respect to ecliptic and
galactic coordinates as well.  All detections recovered as part of
this search were reported to the MPC on 2023 Feb 27 and were published
in MPEC 2023
E12\footnote{\textit{https://www.minorplanetcenter.net/mpec/K23/K23E12.html}}.

\section{Discussion}

The analysis presented here demonstrates that the new software tools
developed for the NEOS mission survey simulator are performing as
expected.  This result establishes a high level of confidence in the
outputs of the simulation in terms of the expected survey
completeness.  While the top-level mission requirement for NEO
Surveyor is to detect two-thirds of the potentially hazardous
asteroids, demonstrating this using the known population requires an
\textit{a priori} knowledge of the population size and properties.
Once NEO Surveyor launches, the NSS code will use the synthetic
population model to evaluate the survey performance in real-time by
tracking the completeness NEO Surveyor would have achieved on the
synthetic population.  Simultaneously, the NSS provides the tools
needed to conduct a full population debiasing at the end of the survey
for the NEOs.  In this way, the survey goals can be doubly-confirmed.

Survey simulation tools like the NSS are necessary for fully
understanding the impacts of design choices on the goals of a survey.
By designing the NSS to be generalizable it is possible to simulate
other surveys and other populations, as we have done here for the
NEOWISE survey.  \citet{grav23} demonstrate a different use case for
the NSS, where they simulated the historical sensitivity of
ground-based surveys to reproduce the NEO discovery rate to the
present day in order to quantify the fraction of objects in our
simulated population that would be expected to be present in the NEO
catalog once NEO Surveyor launches.  In this way we can determine the
overall catalog completeness for PHAs from the worldwide NEO search
efforts.

\section{Conclusions}

The NEO Surveyor mission has developed computational tools to simulate
the observations to be conducted as a way of assessing total expected
performance for the mission.  We have demonstrated here through
comparisons to JPL Horizons, the MPC, and NEOWISE measurements that
the various components of the NSS are functioning as expected and
producing correct positions and fluxes.  This external validation is
critical to establishing the confidence in the expected survey
completeness that NEO Surveyor will deliver.

In addition to this validation analysis, we also use these tools to
search for previously missed detections of NEOs in the NEOWISE data.
We find $21,661$ detections that were previously unreported, which we
have now submitted to the MPC, including data on $1166$ NEOs that
previously had no reported NEOWISE observations\footnote{While
  this manuscript was under review, a similar search of the Year 9
  NEOWISE Reactivation data was completed, resulting in an additional
  2473 NEO detections submitted to the MPC of 439 NEOs, including 170
  objects that had no previous NEOWISE observations, bringing this
  total to 1336.}.  Many of these detections were either extensions
of previously reported tracklets that had been discarded due to
filtering of nearby background objects, while the others were
detections of objects that did not meet the criteria for tracklet
construction used by the NEOWISE automated processing.  As powerful
new software tools like the NSS become available, searches of data
archives like NEOWISE are expected to continue providing important
data that was previously unrecognized for many years to come. This
trend will likely accelerate as the next generation surveys find
orders of magnitude more objects that can be the basis of future
precovery efforts.

\section*{Acknowledgments}
We thank the two anonymous referees for their helpful comments on this
manuscript that improved the text.  This publication makes use of data
products from the Wide-field Infrared Survey Explorer, which is a
joint project of the University of California, Los Angeles, and the
Jet Propulsion Laboratory/California Institute of Technology, funded
by the National Aeronautics and Space Administration.  This
publication also makes use of data products from NEOWISE, which is a
joint project of the University of Arizona and Jet Propulsion
Laboratory/California Institute of Technology, funded by the Planetary
Science Division of the National Aeronautics and Space Administration.
This publication makes use of data products from the NEO Surveyor,
which is a joint project of the University of Arizona and the Jet
Propulsion Laboratory/California Institute of Technology, funded by
the National Aeronautics and Space Administration.  This research has
made use of data and services provided by the International
Astronomical Union's Minor Planet Center.  This research has made use
of the NASA/IPAC Infrared Science Archive, which is operated by the
California Institute of Technology, under contract with the National
Aeronautics and Space Administration.  This research has made use of
the {\it numpy}, {\it scipy}, {\it astropy}, and {\it matplotlib}
Python packages.

Dataset usage:

\begin{itemize}
\item \dataset[WISE All-Sky Single Exposure (L1b) Source Table]{https://www.ipac.caltech.edu/doi/irsa/10.26131/IRSA139}
\item \dataset[WISE All-Sky Single Exposure (L1b) Frame Metadata Table]{https://www.ipac.caltech.edu/doi/irsa/10.26131/IRSA140}
\item \dataset[WISE All-Sky 4-band Single-Exposure Images]{https://www.ipac.caltech.edu/doi/irsa/10.26131/IRSA152}
\item \dataset[WISE 3-Band Cryo Single Exposure (L1b) Source Table]{https://www.ipac.caltech.edu/doi/irsa/10.26131/IRSA127}
\item \dataset[WISE 3-Band Cryo Single Exposure (L1b) Frame Metadata Table]{https://www.ipac.caltech.edu/doi/irsa/10.26131/IRSA128}
\item \dataset[WISE 3-Band Cryo L1b Images]{https://www.ipac.caltech.edu/doi/irsa/10.26131/IRSA149}
\item \dataset[NEOWISE 2-Band Post-Cryo Single Exposure (L1b) Source Table]{https://www.ipac.caltech.edu/doi/irsa/10.26131/IRSA124}
\item \dataset[NEOWISE Post-Cryo L1b Images]{https://www.ipac.caltech.edu/doi/irsa/10.26131/IRSA148}
\item \dataset[NEOWISE-R Single Exposure (L1b) Source Table]{https://www.ipac.caltech.edu/doi/irsa/10.26131/IRSA144}
\item \dataset[NEOWISE-R Single Exposure (L1b) Frame Metadata Table]{https://www.ipac.caltech.edu/doi/irsa/10.26131/IRSA143}
\item \dataset[NEOWISE-R L1b Images]{https://www.ipac.caltech.edu/doi/irsa/10.26131/IRSA147}
\end{itemize}


\begin{thebibliography}{XXX}

\bibitem[Bowell \etal(1989)]{bowell89}
  Bowell E., Hapke B., Domingue D., \etal, 1989, Asteroids II (R. P. Binzel et al., eds), Univ. of Arizona Press, 524.
  
 \bibitem[Cutri \etal(2012)]{cutri12}
 Cutri, R.M., Wright, E., Conrow, T., Bauer, J., \etal, 2012, Explanatory Supplement to the WISE All-Sky Data Release Products, {\it https://wise2.ipac.caltech.edu/docs/release/allsky/expsup}

 \bibitem[Cutri \etal(2015)]{cutri15}
 Cutri, R.M., Mainzer, A., Conrow, T., Masci, F., Bauer, J., \etal, 2015, Explanatory Supplement to the NEOWISE Data Release Products, {\it https://wise2.ipac.caltech.edu/docs/release/neowise/expsup}

\bibitem[Grav \etal(2023)]{grav23}
Grav, T., \etal, 2023, PSJ, submitted.

\bibitem[Harris \etal(1998)]{harris98}
Harris, A.W., 1998, Icarus, 131, 291.

\bibitem[Lebofsky \etal(1978)]{lebofsky78}
  Lebofsky, L. A., Veeder G. J., Lebofsky M. J., and Matson D. L., 1978, Icarus, 35, 336.

\bibitem[Mainzer \etal(2011a)]{mainzer11}
Mainzer, A.K., Bauer, J.M., Grav, T., Masiero, J., \etal, 2011a, ApJ, 731, 53.

\bibitem[Mainzer \etal(2011b)]{mainzer11neatm}
Mainzer, A.K., Grav, T., Masiero, J., \etal, 2011b, ApJ, 736, 100.

\bibitem[Mainzer \etal(2011c)]{mainzer11neo}
Mainzer, A.K., Grav, T., Bauer, J.M., Masiero, J., \etal, 2011c, ApJ, 743, 156.
 
\bibitem[Mainzer \etal(2014a)]{mainzer14neowise}
Mainzer, A.K., Bauer, J., Cutri, R., Grav, T., Masiero, J., \etal, 2014a, ApJ, 792, 30.
 
\bibitem[Mainzer \etal(2014b)]{mainzer14tinyneo}
Mainzer, A.K., Bauer, J., Grav, T., Masiero, J., Cutri, R.,  \etal, 2014b, ApJ, 784, 110.

\bibitem[Mainzer \etal(2019)]{mainzer19}
Mainzer, A.K., Bauer, J., Cutri, R., \etal, 2019,  NASA Planetary Data System. doi:10.26033/18S3-2Z54

\bibitem[Mainzer \etal(2023)]{mainzer23}
Mainzer, A.K., \etal, 2023, PSJ submitted.

\bibitem[Masiero \etal(2011)]{masiero11}
Masiero, J.R, Mainzer, A.K., Grav, T., \etal, 2011, ApJ, 741, 68.
 
\bibitem[Masiero \etal(2017)]{masiero17}
Masiero, J.R, Nugent, C., Mainzer, A.K., Wright, E., Bauer, J., \etal, 2017, AJ, 154, 168.

\bibitem[Masiero \etal(2018)]{masiero18tiny2}
Masiero, J.R, Redwing, E., Mainzer, A.K., Bauer, J.M., Cutri, R.M., \etal, 2018, AJ, 156, 60.

\bibitem[Masiero \etal(2020)]{masiero20tiny3}
Masiero, J.R, Smith, P., Teodoro, L., \etal, 2020, PSJ, 1, 9.
 
\bibitem[Wright(2005)]{wright05}
  Wright, E.L., 2005, New Astronomy Reviews, 49, 407. 

\bibitem[Wright \etal(2010)]{wright10}
Wright, E.L., Eisenhardt, P., Mainzer, A.K., Ressler, M.E., Cutri, R.M., \etal, 2010, AJ, 140, 1868.


\end{thebibliography}
\end{document}